\documentstyle[twoside,fleqn,espcrc2,epsf]{article}
\newcommand{\bi}{\begin{itemize}}
\newcommand{\ei}{\end{itemize}}
\newcommand{\beq}{\begin{equation}}
\newcommand{\eeq}{\end{equation}}

\newcommand{\fig}[1]{Fig.~\ref{#1}}
\newcommand{\tab}[1]{Tab.~\ref{#1}}
\title{Hybrid and Orbitally Excited Mesons in Quenched QCD \thanks{Talk
    presented by P. Lacock}}
\author{ 
        P. ~Lacock$^{\rm a}$, C.~Michael$^{\rm b}$,  
        P. ~Boyle$^{\rm c}$ and P.~Rowland$^{\rm c}$ \\
        (UKQCD Collaboration) \\ [8pt] 
{\rm $^a$}HLRZ, c/o Forschungszentrum J\"ulich, D-52425 J\"ulich,
          and DESY, D-22603 Hamburg, Germany       
{\rm $^b$}Theoretical Physics, Dept. of Mathematical Sciences, 
          University of Liverpool, Liverpool L69 3BX, UK
{\rm $^c$}Physics Department, University of Edinburgh,
           Edinburgh EH9 3JZ, UK\\[8pt]}
\begin{document}
%%%%%%%%%%%%%%%%%%%%%%%%%%%%%%%%%%%%%%%%%%%%%%%%%%%%%%%%%%%%%%%%%%
\begin{abstract}
 
We use lattice methods to evaluate from first principles the spectrum of
hybrid mesons produced by gluonic  excitations  in quenched QCD
with quark masses near the  strange quark mass.   For  the  spin-exotic
mesons with $J^{PC}=1^{-+},\ 0^{+-}$, and $2^{+-}$ which are not present
in the quark model, we determine the lightest state to be $1^{-+}$ with
mass of 2.0(2) GeV.
Results obtained for orbitally excited mesons are also presented. 
\end{abstract}
\maketitle

%%%%%%%%%%%%%%%%%%%%%%%%%%%%%%%%%%%%%%%%%%%%%%%%%%%%%%%%%%%%%%%%%%

\section{INTRODUCTION}

A quantitative study of the hadronic spectrum should include a study of
states which are not in the ground state, i.e. those with excited orbital
angular momentum. Only then can we  get a better insight into the QCD
spectrum. However, a study  of these states requires non-local operators
with the correct symmetries, which can be obtained by either defining
explicit $P$- and/or $D$-wave operators \cite{dh} or     by means of
standard quark propagators combined with some gluonic flux
\cite{prdhyb}. 

Another goal of quantitative studies of QCD is to determine  the
spectrum of gluonic mesons:  the glueballs and hybrid mesons.  Of
special interest are the hybrid meson states with $J^{PC}$ quantum
numbers that are not allowed in the quark model, the so-called
$exotics$. These include the $J^{PC}$ values $1^{-+},\ 0^{+-}$, and
$2^{+-}$. 
 %%Again lattice QCD is capable, from first  principles, of  establishing the
%%masses of these states.

\section{LATTICE OPERATORS}

The group of rotations and inversions of the 3 dimensional
spatial lattice is given by the cubic symmetry group $O_h$ 
(for zero momentum).

In order to construct $lattice$ operators with the desired angular
momentum or gluonic excitation, we have to combine representations of
the `spin' cubic group (coming from the quark spinors) with the
`orbital' cubic group (coming from the spatial paths). Following
\cite{prdhyb} we study non-local gluonic fields in specific
representations of the lattice rotation group.

In order to construct optimal lattice operators, that is with optimal
signal to noise ratio, we need operators with a given $O_h$ representation
at the source and/or sink.

The operators used here involve a `white' source, i.e. a source which, in
principle, gets contributions from all quantum numbers, while the sink
operators are constructed to project out the desired symmetry \cite{prdhyb}. 

We work on a $16^3 \times 48$ lattice at $\beta$ = 6.0 and use a tadpole
improved clover action with $\kappa$ = 0.137 and $c_{sw}$ = 1.4785. This
value of the hopping parameter corresponds to a quark mass around the 
strange quark mass. We have generated a set of 350 local propagators
consisting of sources for the quark at (0,0,0,0) and at (0,0,6,0) for
the anti-quark. At the source the propagators are connected by a path
$\cal P$ consisting of  fuzzed gluon links. The path $\cal P$ is chosen
in order to provide the desired angular momentum (in the case of the
$L$-excited mesons) or the gluon excitation (for the hybrid mesons). The
resulting hadronic correlations in both cases are therefore also by
definition gauge invariant. Fuzzed links are used to improve the overlap
with the ground state \cite{prdhyb,plhyb}.

An alternative procedure to the method used here is to
study hybrid mesons using the continuum symmetries \cite{hyb}.

\begin{figure}[htb]
\vspace{-2.3cm}
%\leavevmode
\epsfxsize=9.5cm\epsfbox{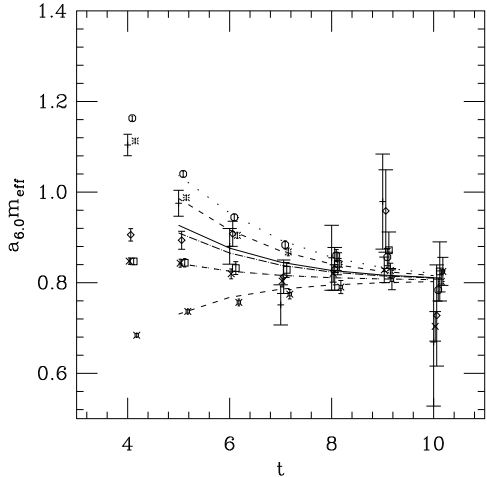}
\vskip -1cm
\hskip -6mm
 \caption{ The lattice effective mass for the $1^{++}$ meson vs. time $t$.
 For a discussion of the different operators used, see ref. \cite{prdhyb}.
\label{A1}}
\vspace{-0.7cm}
\end{figure}

\section{$L$-EXCITED MESONS}

$L$-excited mesons can be studied by choosing $\cal P$ to be the
straight product of  fuzzed links connecting the quark and anti-quark at
the source and sink. At the source both the direction (here the \^z or 3
direction) and the length (=6) are fixed. At the sink, on the other
hand, we have three possible spatial directions, and the length $R$ can
also be varied.

We perform correlated 2-state fits to the effective mass and use as many
operators (i.e. different spatial link combinations and  choices of $R$)
to constrain the fits as much as possible. As a typical example, we show
the results for the $A_1$ meson in \fig{A1}.

We are also able to determine the hyperfine splitting among the four
states in the $P-$wave multiplet, finding that  the singlet state has
the lowest mass, while the other three members have masses which are
degenerate within the statistical errors. This result, reported in
\cite{prdhyb}, is now confirmed using higher statistics.

\section{HYBRID MESONS}

Hybrid mesons are by definition mesons with an excited gluonic
component.  From studies of static quarks it was found that the lowest
order lying  hybrid states have  colour flux from the quark to
anti-quark excited in a transverse spatial plane~\cite{cmper}.  This can
be achieved by the choice of U-shaped paths of fuzzed links at the
source and sink. Here we use the same construction for lighter quark
masses \cite{prdhyb}.

The lowest lying gluonic excitations have  spatial symmetries
corresponding to $L^{PC}$ = $1^{+-}$ and $L^{PC}$ = $1^{-+}$. Combining
these  with the $q \bar q$ spin representations, we obtain a range
of possible $J^{PC}$ values which include the spin-exotic quantum
numbers  $J^{PC}=1^{-+},\ 0^{+-}$, and $2^{+-}$ which are not present 
in the quark model. 

\begin{figure}[htb]
\vspace{-3.0cm}
\leavevmode
\epsfxsize=9.5cm\epsfbox{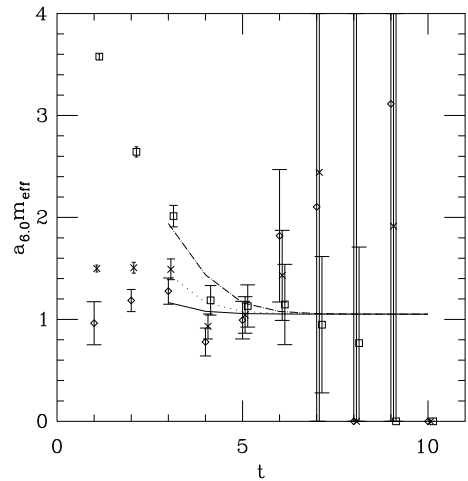}
\vskip -1cm
%\normalpicture{hybu1d.ps}
 \caption{ The lattice effective mass for the $J^{PC}=1^{-+}$  exotic
hybrid meson vs $t$. The source used is a U-shaped path of size  $6
\times 6$, while the sinks are combinations of U-shaped paths of size $6
\times 6$ ($\diamond$), $3 \times 3$ ($\Box$) and $1 \times 1$ (fancy
cross).
 \label{exot1}}
\vspace{-8mm}
\end{figure}

In our simulation the source necessarily again has longitudinal length
($r$)=6 fixed, while the transverse length ($d$) is free. At the sink
one can vary both ($R$ and $D$ respectively). From experience we found
that the optimum  choice of operators has ($r,d$)=(6,6) at the source,
while at the sink we use ($R,D$) = (1,1), (3,3) and (6,6). The choice of
(6,6) at both source and sink moreover gives an upper bound on the
ground state mass.

The results obtained from the correlated 2-state
fits are listed for the exotic hybrid mesons
in \tab{RES} and shown for the $1^{-+}$ exotic in \fig{exot1} \cite{plhyb}.  
Our results are consistent with the assumption that the
$1^{-+}$ state is the lightest \cite{close}. Taking account of statistical
correlations, we find that the $0^{+-}$ state is heavier  than the
$1^{-+}$ state with a significance of 1 standard deviation.

\begin{table}[htb]
\vspace{-5mm}
\footnotesize
\begin{tabular}{lclll}
\hline
meson&  $J^{PC} $  & mass& $M/M_V$&$M_s/M(\phi)$\\
 multiplet &      &     $Ma$&  &\\
\hline
$\hat{\rho}$&$  1^{-+},\ 3^{-+} $ &0.95(7)&1.76(13) &1.95(13)\\
\^a$_0$&$ 0^{+-},\ 4^{+-}$  & 1.05(7)& 1.94(13)&2.16(13)\\
\^a$_2$&$  2^{+-},\ 4^{+-} $   & 1.26(13)& 2.33(24)&2.62(24)\\
\hline
\vspace{8pt}
\end{tabular}
 \caption{The masses of the exotic hybrid mesons. The $J^{PC}$ values
are the lowest two values allowed by the lattice cubic symmetry.   The
masses are given in lattice units and as a ratio to the vector meson
mass $m_V$. The mass ratio  appropriate to   $s \bar{s}$ states is given
in the last column \cite{plhyb}.}
 \label{RES}
\vspace{-5mm}
\end{table}

In \tab{RES} we also list  our lattice results for the hybrid masses as
a ratio to  the vector meson mass we find on the same lattices, namely
$M_V a=0.54(1)$. Since we only have results at one quark mass value and
lattice spacing, we determine the value of the exotic mesons by  using
the experimental $\phi$ meson mass to set the scale. Our result for the
lightest $s \bar{s}$ hybrid meson then is 1.99(13) GeV. Taking into
account some of the  possible systematic errors  (lattice size effects,
extrapolation uncertainties etc.), we round this prediction to 2.0(2)
GeV. This value is in agreement with the value obtained by the MILC
Collaboration \cite{hyb}.

\begin{figure}[htb]
\vspace{-2.4cm}
\leavevmode
\epsfxsize=9.5cm\epsfbox{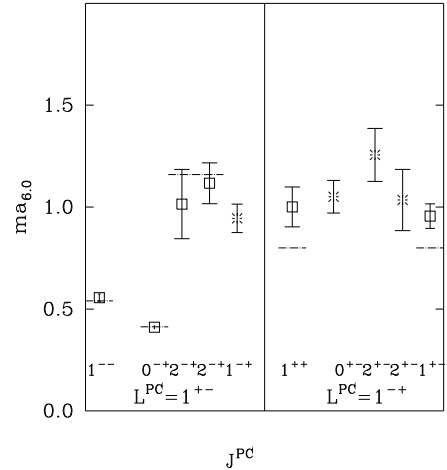}
\vskip -1cm
\caption{
 Ordering of the hybrid meson levels. The bursts denote the $J^{PC}$
exotic states.
 \label{order}}
\vspace{-3mm}
\end{figure}

\begin{table}[htb]
\vspace{-5mm}
\footnotesize
\begin{tabular}{lll}
\hline
meson&  $J^{PC} $  & mass($Ma$) \\ 
\hline
$\rho$&$  1^{--} $ &0.541(2)\\
$\pi$&$  0^{-+} $ & 0.413(1)\\
$B   $&$  1^{+-} $ & 0.801(20)\\
$A_1  $&$  1^{++} $ & 0.808(20)\\
$ \pi_2  $&$  2^{-+} $ & 1.165(60)\\
\hline
\vspace{8pt}
\end{tabular}
 \caption{The masses of the $L$-excited mesons shown in \fig{order}.} 
\label{RES1}
\vspace{-8mm}
\end{table}

In \fig{order} we show the ordering of the hybrid meson levels. The
dashed lines represent $L$-excited states for the same full set of
configurations which are listed in \tab{RES1}. The strong mixing of the
hybrid mesons and standard mesons (i.e. with no gluonic excitation) with
the same $J^{PC}$ values even in the quenched approximation is apparent
for the pseudoscalar and vector meson cases.

In principle one can also construct hybrid mesons using local gluonic
fluxes, i.e. gluon loops starting and ending at the same site. Except
for the lowest lying exotic hybrid state, the results obtained are 
found to be more noisy~\cite{plhyb}, suggesting that local gluonic
operators might not be an ideal choice to provide the gluonic
excitations of interest.

%%\section*{Acknowledgements}
We acknowledge support from EPSRC grant GR/K/41663 to the UKQCD
Collaboration.  
%%%%%%%%%%%%%%%%%%%%%%%%%%%%%%%%%%%%%%%%%%%%%%%%%%%%%%%%%%%%%%%%%%


\vspace{-2mm}
\begin{thebibliography}{99}
\frenchspacing
%

%\bibitem{chris}  see e.g.
%C. Michael, Acta Phys. Pol. B {\bf 21} (1990) 119

\bibitem{dh}   T.A. DeGrand and M.W. Hecht, 
Phys. Rev. D {\bf 46} (1992) 3937.

\bibitem{prdhyb}   UKQCD Collaboration, P. Lacock et al., 
Phys. Rev. D {\bf 54} (1996) 6997.

\bibitem{plhyb}   UKQCD Collaboration, P. Lacock et al., 
Phys. Lett B {\bf 401} (1997) 308.
 
\bibitem{hyb} C. Bernard et al., these proceedings.
 
\bibitem{cmper}S.J. Perantonis and C. Michael, Nucl. Phys. B {\bf 347} (1990) 854.
 
 
\bibitem{close} F.E. Close, Proc.~Hadron Conference, Manchester 1995;
hep-ph/9509245.
 
 
%
\end{thebibliography}
\end{document}